\documentclass[twocolumn,showpacs,preprintnumbers,amsmath,amssymb]{revtex4}
\usepackage{graphicx}% Include figure files
\usepackage{rotating}% rotating figure files
\usepackage{dcolumn}% Align table columns on decimal point
\usepackage{bm}% bold math

\begin{document}
%\preprint{APS/123-QED}
\date{\today}
%%%%%%%%%%%%%%%%%%%%%%%%%%%%%%%%%%%%%%%%%%%%%%%%%%%%%%%%%%%%%%%%%%%%%%%%%%%%%%%%%
%                    h-BN article                                               %
%                                                                               %
%          written by B. Arnaud.                                                %
%          GMCM Rennes France                                                   %
%%%%%%%%%%%%%%%%%%%%%%%%%%%%%%%%%%%%%%%%%%%%%%%%%%%%%%%%%%%%%%%%%%%%%%%%%%%%%%%%%
\title{Huge excitonic effects in layered hexagonal boron nitride}
\author{B.~Arnaud$^{1}$, S.~Leb\`egue$^{2}$,  P.~Rabiller$^{1}$, and M.~Alouani$^{3}$}
\affiliation{$^{1}$Groupe Mati\`ere condens\'ee et Mat\'eriaux (GMCM),
Campus de Beaulieu - Bat 11 A, 35042 Rennes Cedex, France, EU}
\affiliation{$^{2}$Department of Physics, Uppsala University, SE-75121 Uppsala, Sweden, EU}
\affiliation{$^{3}$Institut de Physique et de Chimie des Mat\'eriaux de Strasbourg (IPCMS),
UMR 7504 CNRS-ULP, 23 rue du Loess, 67034 Strasbourg, France, EU}
\date{\today}% It is always \today, today,

\begin{abstract}
The calculated quasiparticle band structure of bulk hexagonal boron nitride 
using the all-electron GW approximation shows that this
compound is an indirect-band-gap semiconductor. 
The solution of the Bethe-Salpeter equation for the
electron-hole two-particle Green function has been  used to compute its 
optical spectra and the results are found in
excellent  agreement with available experimental data. A detailed
analysis is made for the excitonic structures within the band gap and found
that the excitons belong to the Frenkel class and are tightly confined within
the layers. The calculated exciton binding energy is much larger
than that obtained by Watanabe {\it et al}\cite{watanabe_taniguchi} 
using a Wannier model to interpret their experimental results and assuming 
that h-BN is a direct-band-gap semiconductor.  
	 
\end{abstract}
\maketitle
%%%%%%%%%%%%%%%%%%%%%%%%%%%%%%%%%%%%%%%%%%%%%%%%%%%%%%%%%%
%%%%%%%%%%%%%%%%%%%       INTRO    %%%%%%%%%%%%%%%%%%%%%%%
%%%%%%%%%%%%%%%%%%%%%%%%%%%%%%%%%%%%%%%%%%%%%%%%%%%%%%%%%%
Hexagonal boron nitride (h-BN) is one of the most anisotropic layer compounds
and represents an interesting quasi-two-dimensional insulator analog to
semi-metallic graphite. Due to its high thermal stability, h-BN is a widely
used material in vacuum technology. It has been employed for microelectronic
devices\cite{pauli}, for x-ray lithography masks\cite{dana}, and as a
wear-resistant lubricant\cite{miyoshi}. The interest in h-BN has been renewed
by the possibility of  preparing  boron nitride nanotubes that are far more
resistant to oxidation than carbon nanotubes and therefore suited to high
temperature applications. As their band gaps are predicted to be weakly
dependent on helicity and tube diameter\cite{rubio_corkill, blase_rubio},
unlike tubes made of carbon, it has the potential of revolutionizing the
electronics industry.  To understand the quasiparticle  properties of BN
nanotubes one has first to have a   complete understanding of the electronic and
optical properties of  bulk h-BN since BN nanotubes can be viewed as a rolled
up BN sheets.

Despite the large number of experiments\cite{tegeler, zunger_katzir, hoffman,
mamy, watanabe_taniguchi} devoted to the study of the electronic properties of
bulk h-BN, both the direct and the indirect band-gap are not yet accurately known,
i.e., values ranging from 3.2 to
5.97 eV have been reported in the literature. In particular, 
the latest results were obtained by Watanabe
{\it et al}\cite{watanabe_taniguchi},  who showed that this material is very
promising for ultraviolet laser devices and assumed, based on the strong luminescence  
peak observed at 5.765 eV, that it is a direct-band-gap semiconductor. In addition,
they inferred a gap of 5.971 eV in contradiction with
our results and other GW quasiparticle 
calculations\cite{blase, capellini_satta, capellini_fiorentini}.

In this Letter, we use our all-electron GW approximation\cite{brice} to study
for the first time the optical properties of bulk h-BN including electron-hole
interaction effects and show that the previous experimental interpretation of
the excitonic structure in the band gap is not correct.  To perform  this study
we used  state of the art methods, where both the  self-energy effects  and the
solution of the Bethe-Salpeter equation\cite{onida_reining} to compute the
optical spectra from first principles\cite{arnaud_2001}.  In particular, our
study confirms  that this material is an indirect-band-gap semiconductor in
agreement with previous GW
calculation\cite{blase,capellini_satta,capellini_fiorentini}, and that the
interpretation of the optical spectra of h-BN, upon which  the band gap  of
Watanabe {\it et al}\cite{watanabe_taniguchi} is based, is very intricate as
they are dominated by strong electron-hole interaction effects.  We show that
theoretical optical spectra are  reliable  and   are  in excellent agreement
with experiment\cite{mamy}.  In addition, we show that the exciton in the
band gap belongs to a Frenkel class instead of a Wannier class as proposed by
Watanabe {\it et al}\cite{watanabe_taniguchi} from the analysis of 
their experimental data, and we found that its  binding energy is huge 
compared  to theirs.

%%%%%%%%%%%%%%%%%%%%%%%%%%%%%%%%%%%%%%%%%%%%%%%%%%%%%%%%%%%%%%%%%%%%%%%%%%%%%%
%%%%%%%%%%%%%%%%%%%%%%%%%%%%%%%%%%%%%%%%%%%%%%%%%%%%%%%%%%%%%%%%%%%%%%%%%%%%%%
%%%%%%%%%%%%%%%%%%%%%%%%%%%%%%%%%%%%%%%%%%%%%%%%%%%%%%%%%%%%%%%%%%%%%%%%%%%%%%
%\section{Quasiparticle properties}
%{\sl Quasiparticle properties:}
The crystalline structure of  h-BN is hexagonal and  has the D$_{6h}^{4}$ group
symmetry. It consists of hexagonal graphite-like sheets but with an
ABAB$\cdots$ stacking with boron atoms in layer A found directly below nitrogen
atoms in layer B. The ground state properties of h-BN are obtained by means of
the all-electron Projector-Augmented-Wave method (PAW)\cite{Bloechl} from
density-functional theory (DFT) in the local-density approximation
(LDA)\cite{perdew}, yielding LDA band-structure energies, and wave functions needed
to calculate all quantities occurring in the GW self-energy operator and
the electron-hole interaction. The experimental lattice
parameters\cite{solozhenko} (a=4.73 a.u and c/a=2.66) are used to perform all
calculations.

% \begin{widetext}
% \begin{equation}
% \left[-\frac{\nabla^2_{\bf r}}{2}+V_{external}({\bf r})+V_{Hartree}({\bf r}) \right]
% \Psi_{n{\bf k}}({\bf r})
% + \int d^3r^{\prime}
% \Sigma({\bf r},{\bf r}^{\prime}, \epsilon_{n{\bf k}}^{qp})
% \Psi_{n{\bf k}}({\bf r}^{\prime})
% = \epsilon_{n{\bf k}}^{qp}\Psi_{n{\bf k}}({\bf r}),
% \label{qp_equation}
% \end{equation}
% \end{widetext}
The quasiparticle energies are determined by solving the QP equation,
where the self-energy operator $\Sigma$ is written down in Hedin's 
GW approximation \cite{hedin} which includes dynamic polarization
in the random-phase approximation (RPA). In this case $\Sigma$ can be
written as the product of the Green's function and the dynamically 
screened Coulomb interaction within the RPA.
% \begin{equation}
% \Sigma ({\bf r}, {\bf r}^\prime , \omega)=
% \frac{i}{2\pi} \int e^{i\delta \omega^\prime}
% G({\bf r}, {\bf r}^\prime , \omega+\omega^\prime) W({\bf r}, {\bf r}^\prime ,
% \omega^\prime) {\rm d}\omega^\prime,
% \label{selfenergy}
% \end{equation}
% where $W=\epsilon^{-1}v$ is the dynamically screened interaction, $v$ is the
% bare Coulomb potential and $\epsilon^{-1}$ is the inverse of the dielectric
% matrix computed within RPA.  
In this work, a plasmon pole model\cite{horsch} is
used to mimic the frequency dependence of the dielectric matrix in order to
reduce the computational cost of the GW calculations. More details about our
implementation are given elsewhere\cite{brice}.
%\begin{figure}[h!]
\begin{figure}
\includegraphics[width=8.5cm]{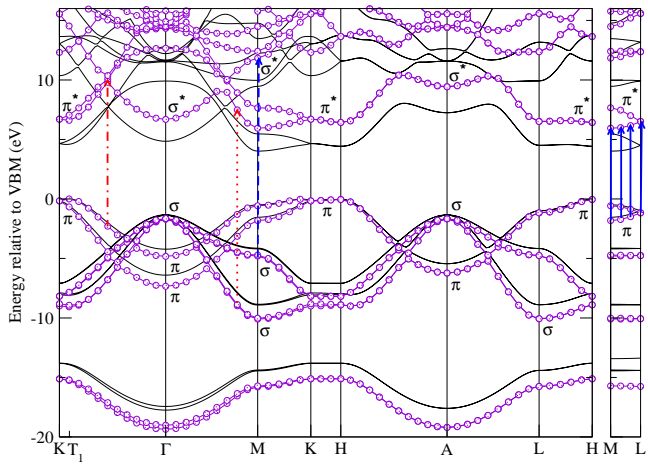}
%\centerline{ \includegraphics*[angle=0,width=\textwidth]{fig1.eps}}
\caption{\label{fig:band_structure}
Calculated electronic band structure along high-symmetry lines for bulk
hexagonal boron nitride.  The thin solid lines display the LDA results
while the solid lines with open circles
show the GWA results. The energy scale is relative
to the top of the valence band maximum (VBM) located at the T$_1$ point near K
along the $\Gamma$-K high symmetry direction.  } \end{figure}

In Fig. \ref{fig:band_structure}, both the calculated LDA (thin solid lines)
and GW (solid lines with open circles) band structures are plotted along the
high symmetry directions.  Within the LDA, we found an indirect gap of 4.02 eV
between the bottom of the conduction band at the $M$ point and the top of the valence
band near $K$ ($T_1$ point).  The minimum direct gap is located at the $H$ point and is
4.46 eV.  Our results are in excellent agreement with other LDA calculations
\cite{blase, xu_ching, furthmuller}.  The main effect of the self-energy
correction is to increase the gap from 4.02 eV (LDA) to 5.95 eV and the minimum
direct gap from 4.46 eV (LDA) to 6.47 eV. Our GW results are in good agreement
with the calculations of Capellini {\it et al}\cite{capellini_fiorentini} who found
values 6.04 and 6.66 eV, respectively.  Figure
\ref{fig:band_structure} shows also that the GW corrections to the LDA are
strongly dependent on the degree of orbital localization and that the overall
corrections can not be reproduced by a rigid band shift of the conduction states
with respect to the top of the valence states.  Such a behavior has
already been reported for layered compounds like graphite\cite{strocov} or LiBC
\cite{lebegue_libc}.

The wide range of experimental 
values\cite{tegeler, zunger_katzir, hoffman, mamy, watanabe_taniguchi} 
reported for the band gap (between 3.2 and 5.97 eV)
may be due, in large part, to the difficulty to synthesize high quality h-BN
crystals or to the difficulty to extract the band gaps from experiment.
The most accurate results have been obtained by Watanabe {\it et
al.}\cite{watanabe_taniguchi} who resolved the excitonic structure within the
gap at very low temperature (8 K) and deduced from the identification of a
plateau in their absorption spectrum that the QP gap is 5.971 eV. Notice that the
inferred band gap is very close to our calculated indirect band gap. It's worth 
pointing out that the interpretation of the optical spectra of non-conducting
solids is very intricate because of strong electron-hole correlation effects.

%This 
%is puzzling because Watanabe {\it et al}\cite{watanabe_taniguchi} claimed
%that h-BN is a direct band gap insulator.

%%%%%%%%%%%%%%%%%%%%%%%%%%%%%%%%%%%%%%%%%%%%%%%%%%%%%%%%%%%%%%%%%%
%%%%%%%%%%%%%%%%%%%%%%%%%%%%%%%%%%%%%%%%%%%%%%%%%%%%%%%%%%%%%%%%%%
%%%%%%%%%%%%%%%%%%%%%%%%%%%%%%%%%%%%%%%%%%%%%%%%%%%%%%%%%%%%%%%%%%
%\section{Electron-hole excitations and optical spectra}
%{\sl Electron-hole excitations and optical spectra:}
In the absence of exciton-phonon coupling, emission or absorption of a
photon involves excited states $|\Psi^{\lambda}_{{\bf q}}\rangle$ of total
wavevector ${\bf q}$ corresponding to the photon momentum, hereafter approximated as
${\bf q}={\bf 0}$. 
Due to electron-electron interaction, these excited states
can be viewed as a linear combination of quasielectron-quasihole pairs: 
\begin{equation}
|\Psi^{\lambda}_{{\bf q}={\bf 0}}\rangle=\sum_{{\bf k},c,v} A^{\lambda}_{vc{\bf k}}
a^+_{c{\bf k}} a_{v{\bf k}}|0\rangle \end{equation} 
where $a^+_{c{\bf k}}$ creates a quasielectron and $a_{v{\bf k}}$ a quasihole
in the many-body ground state $|0\rangle$. The
electron-hole amplitudes $A^{\lambda}_{vc{\bf k}}$ and the excitation
energies $E^{\lambda}$ are obtained by solving an effective
two-particle Schr\"odinger equation, which originates from the
Bethe-Salpeter equation\cite{strinati} and provides the key ingredients to
calculate the imaginary part of the dielectric function including both local field (LF) and excitonic
effects\cite{arnaud_2001, arnaud_2005}.

As the screening controls the strength of the electron-hole interaction, we
inverted directly the dielectric function at the RPA level to obtain the static
dielectric function $\epsilon_{\infty}({\bf q} \to 0)$. The LDA,  without LF
effects,  produces values of 4.67 and 2.98 for ${\bf E}\perp {\bf c}$ and ${\bf
E}\parallel{\bf c}$, respectively, where ${\bf E}$ is the electric field
vector.  As expected, the LF effects push the oscillator strength towards
higher energies\cite{marinopoulos} reducing the static dielectric
function  values to 4.40 and 2.53, respectively. Notice that the LF have a
stronger effect for the out of plane polarization, i.e., ${\bf E}\parallel{\bf
c}$. This is  due to the rather large inhomogeneity of the h-BN charge density
along the ${\bf c}$ direction.  These results show that the screening is rather
anisotropic and that excitonic effects are expected to be important, i.e.,  the
smaller dielectric constant,   the larger 
the excitonic effects due to the poor screening of the electron-hole interaction.

%\begin{figure}[h!]
\begin{figure}
\includegraphics[width=8.5cm]{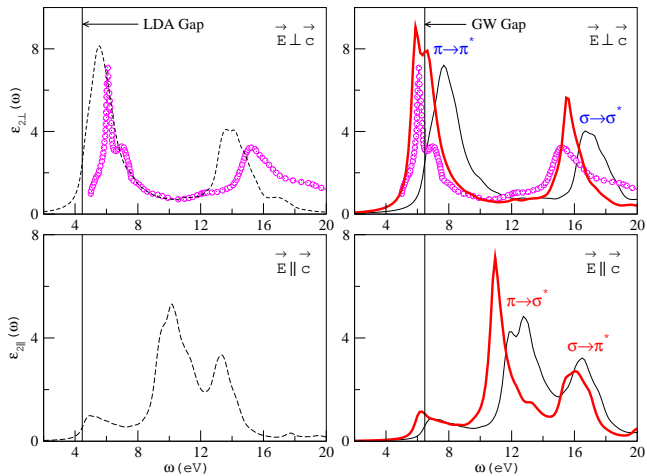}	
%%\centerline{ \includegraphics*[angle=0,width=\textwidth]{fig2.eps}}
\caption{\label{fig:dielectric_function}
Calculated imaginary part of the dielectric function of bulk hexagonal boron
nitride with (thick solid line) and without (dashed line and thin solid line)
excitonic effects for light polarized perpendicular (in-plane polarization) and
parallel to the {\bf{ c}}-axis versus photon energy compared with the
experimental data (empty circles) of Ref.\cite{mamy}. The dashed
lines and the thin solid lines are calculated using the LDA 
and the GW eigenvalues, respectively.
}
\end{figure}

The calculated optical absorption spectra for h-BN with and without excitonic
effects are shown in Fig. \ref{fig:dielectric_function} and are compared to
the experimental results\cite{mamy} for in-plane polarization (${\bf E}\perp {\bf
c}$). The LDA-RPA (dashed line) and the GW-RPA (thin solid line)  dielectric
functions are red-shifted and blue-shifted with respect to experiment, respectively. 
Notice that the GW-RPA spectra can not be obtained by simply shifting
the LDA-RPA spectra towards higher energies because the corrections to the LDA
eigenvalues are far from being uniform across the BZ. The GW-RPA spectrum for
the in-plane polarization exhibits two prominent structures located at 7.65 eV and
16.65 eV, respectively. The first structure originates from the  $\pi\rightarrow\pi^*$
interband transitions between two parallel bands along the M-L high symmetry direction 
and is  denoted by solid
arrows in Fig. \ref{fig:band_structure}. For such transitions, the charge
densities for the valence and the conduction states both have a $p_z$ character and are
located on nitrogen and boron atoms, respectively. The second peak originates
from $\sigma\rightarrow\sigma^{*}$ interband transitions. One of the corresponding
interband transition,  occurring at the M point,  is depicted by a dashed arrow on
Fig. \ref{fig:band_structure}.  The GW-RPA spectrum for out-of-plane polarization
(${\bf E}\parallel{\bf c}$) displays three major structures located at 12 eV,
12.75 eV and 16.5 eV and  originate from $\pi\rightarrow\sigma^{*}$ or
$\sigma\rightarrow\pi^{*}$ interband transitions.  The dot-dashed and dotted arrows on
Fig.  \ref{fig:band_structure} represent one of the interband transitions
contributing to the peaks located at 12.75 eV and 16.5 eV, respectively. 

The assignment of the optical transitions for both polarization directions can
be understood by invoking symmetry arguments. Since there is a mirror
plane within each layer, it's possible to separate bands into states of even
($\sigma$) and odd ($\pi$) parity. Therefore, only transitions between bands of
the same symmetry ($\pi\rightarrow\pi^*$ and $\sigma\rightarrow\sigma^{*}$) are
allowed for in-plane polarization (${\bf E}\perp {\bf c}$), while only
transitions between bands of different parity ($\pi\rightarrow\sigma^{*}$ and
$\sigma\rightarrow\pi^{*}$) are allowed for out-of-plane polarization.
%(${\bf E}\parallel{\bf c}$). 
Notice that this selection rule is only valid for ${\bf k}$
points whose component along the ${\bf c}^*$ axis is strictly zero. Thereby, we
can explain the rather small but non vanishing oscillator strength below 10 eV
for the out-of-plane polarization.
%\begin{figure}[h!]
\begin{figure}
 \includegraphics[width=10.0cm]{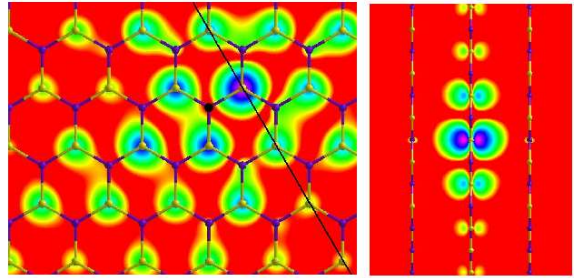}
%\centerline{ \includegraphics*[angle=0,width=\textwidth]{fig3.eps}}
\caption{\label{fig:excitonic_function}
Two dimensional projections of the probability density $|\Psi^{\lambda}({\bf
r_h}, {\bf r_e})|^2$ for the exciton state at 5.780 eV.  The left panel shows
the distribution of the electron relative to the hole located slightly above a
nitrogen atom (black circle) in a plane containing the hole and perpendicular to the
{\bf{ c}} axis. The right panel shows the distribution of the electron relative
to the hole in a plane parallel to the {\bf{ c}} axis passing through the line
shown in the left panel.  } 
\end{figure}
Figure \ref{fig:dielectric_function} also shows the calculated spectra
including excitonic effects (thick solid line). The electron-hole attraction shifts
the oscillator strength towards lower photon energies by about 1 eV, bringing the
calculated spectra in better agreement with experiment.  A bound exciton shows
up at 5.85 eV in the calculated spectra while a prominent peak occurs at 6.08
eV in the experimental spectrum. It is interesting to point out
that a peak around 6.15 eV, characteristic
of h-BN, has been found in the absorption spectrum of single-wall boron nitride
nanotubes\cite{lauret}.  A detailed analysis of our calculated peak shows that it's made
of four excitonic lines located at 5.75 , 5.78 , 5.82 and 5.85 eV. Each of this
excitonic state is built from a coherent superposition of
quasielectron-quasihole pairs corresponding to $\pi\rightarrow\pi^*$
transitions between the two last valence bands and the two first conduction
bands for ${\bf k}$ points laying in the HLMK plane. The ${\bf k}$ points
which contribute the most to these excitonic lines are distributed along directions parallel to
the ${\bf c}^*$ axis and in the direction HK, where the bands are nearly parallel (see
Fig. \ref{fig:band_structure}). 
Notice that these excitonic peaks are only visible for in-plane
polarization because bands of the same parity ($\pi\rightarrow\pi^*$) are
involved in these excited states. Again, it's worth comparing our results with
those of Watanabe {\it et al.}\cite{watanabe_taniguchi} who found four excitonic
lines located at 5.822, 5.945, 5.962 and 5.968 eV, respectively. These lines 
were attributed
to a series of s-like excitons ($n=1,\cdots,4$) by using a Wannier model and
estimating the band gap to be 5.971 eV. The overall agreement with our results
is pretty good when we shift our calculated excitonic lines by 0.07 eV towards
higher energy.  Nevertheless, we do not reproduce the experimental spacing
between the different lines. Such a difference might be due to
the approximations made to solve the Bethe-Salpeter
equation. Regarding the interpretation of the excitonic structures inside the
gap, our results strongly disagree with those of Watanabe {\it et al} who
inferred a binding energy of 0.149 eV from their optical absorption spectra.
Indeed, our results show that the binding energy is much larger,  0.72 eV,  and 
that excitons in h-BN belong to the Frenkel and not to the Wannier class.

To address the localization of the excitons for h-BN, we calculate the exciton
probability $|\Psi^{\lambda}({\bf r_h}, {\bf r_e})|^2$ for a given hole
position ${\bf r_h}$ and show the distribution of the electron relative to the
hole for the excited state at 5.78 eV.  Figure \ref{fig:excitonic_function}
displays such a distribution for a hole localized slightly above a nitrogen atom. The
left panel shows that the  electron is delocalized on boron atoms and that the
probability of finding the electron on a boron atom is maximum for the nearest
neighbors of the nitrogen atom where the hole is localized.  The right panel shows
that the electron is confined within the layer where the hole is located, which is
expected because the screening is rather poor along the ${\bf c}$ direction.
The distribution of the electron relative to the hole also reflects the orbital
character of the conduction states which have a p$_z$ character on B atoms.
Similar plots have been obtained for other bound excitons, confirming that the
Frenkel exciton-like picture is valid.

%%%%%%%%%%%%%%%%%%%%%%%%%%%%%%%%%%%%%%%%%%%%%%%%%%%%%%%%%%%%%%%%%%
%%%%%%%%%%%%%%%%%%%%%%%%%%%%%%%%%%%%%%%%%%%%%%%%%%%%%%%%%%%%%%%%%%
%%%%%%%%%%%%%%%%%%%%%%%%%%%%%%%%%%%%%%%%%%%%%%%%%%%%%%%%%%%%%%%%%%
%%%%%%%%%%%%%%%%%%%%%%%%%%%%%%%%%%%%%%%%%%%%%%%%%%%%%%%%%%%%%%%%%%
%\section{Conclusion}
In this letter, results concerning the excited states of h-BN have been
presented. By calculating the quasiparticle band structure within the GW
approximation, we found that h-BN is an indirect gap semiconductor with a band
gap energy of 5.95 eV in surprisingly close agreement with the supposed direct
gap of 5.971 eV reported by Watanabe {\it et al}\cite{watanabe_taniguchi} while
our calculated minimum direct band gap is 6.47 eV. We solved  the
Bethe-Salpeter equation to calculate the optical spectra and obtained a good
agreement with the experimental results of Ref.\cite{mamy} for in-plane
polarization up to 20 eV. A detailed analysis of the four excitonic structures
within the gap shows that these structures are related to $\pi\rightarrow\pi^*$
transitions across the gap. In addition, we deduced a binding energy of 0.72 eV
in strong disagreement with the value of 0.149 eV found in Ref.
\cite{watanabe_taniguchi}. On the one hand, the large excitonic binding energy
($\gg$ kT) explains the unusual strong luminescence peak for an indirect
semiconductor observed by Watanabe {\it et al} at 5.765 eV, and on the other hand
it explains why this peak does not show noticeable temperature dependence.
Finally, we believe that this work is a first step towards a better
understanding of large excitonic effects reported for single-wall boron nitride
nanotubes\cite{lauret}.

%\begin{acknowledgments}
Supercomputer time was provided by the CINES (project gmg2309) on the IBM SP4. We would 
like to acknowledge partial support from the GDR-DFT. 
%\end{acknowledgments}

\end{document}